\documentclass[aps,prl,twocolumn,superscriptaddress,nobibnotes,longbibliography,10pt]{revtex4-2}
\usepackage{graphicx}% Include figure files
\usepackage{dcolumn}% Align table columns on decimal point
\usepackage{bm}% bold math
\usepackage[utf8]{inputenc}
\usepackage[T1]{fontenc}
\usepackage{mathptmx}
\usepackage{etoolbox}
\usepackage{xcolor}
\usepackage{float}
\usepackage{amsmath}
\usepackage{amssymb}
\usepackage{hyperref}
\usepackage{comment}
\usepackage{todonotes}
\usepackage{xr-hyper}
\usepackage{cleveref}
\usepackage{braket}

\begin{document}

%\title{Velocimetry with stopped light at room temperature}
\title{Controlled Displacement of Stored Light at Room Temperature}

\author{Arash Ahmadi}
\email{arash.ahmadi@physik.hu-berlin.de}
\affiliation{Institut für Physik and Center for the Science of Materials Berlin (CSMB), Humboldt-Universität zu Berlin, Newtonstr. 15, Berlin 12489, Germany}

\author{Ya{\u{g}}{\i}z Murat}
\affiliation{Institut für Physik and Center for the Science of Materials Berlin (CSMB), Humboldt-Universität zu Berlin, Newtonstr. 15, Berlin 12489, Germany}

\author{Pei-Chen Kuan}
\affiliation{Department of Physics, National Cheng Kung University, Tainan 70101, Taiwan}

\author{Mustafa Gündo{\u{g}}an}
\email{mustafa.guendogan@physik.hu-berlin.de}
\affiliation{Institut für Physik and Center for the Science of Materials Berlin (CSMB), Humboldt-Universität zu Berlin, Newtonstr. 15, Berlin 12489, Germany}

\author{Markus Krutzik}
\affiliation{Institut für Physik and Center for the Science of Materials Berlin (CSMB), Humboldt-Universität zu Berlin, Newtonstr. 15, Berlin 12489, Germany}
\affiliation{Ferdinand-Braun-Institut (FBH), Gustav-Kirchoff-Str. 4, Berlin 12489, Germany}

\date{\today}

\begin{abstract}
We report the demonstration of spatially translating a stored optical pulse at room temperature over distances exceeding one optical wavelength. By implementing an interferometric scheme, we further measure the average speed of this linear translation, thus harnessing a stopped-light experiment for a sensing application. This work extends the use of quantum memories beyond quantum communication and information contexts, opening a pathway to novel methods of velocity measurements with high sensitivity. \end{abstract}

\maketitle

Quantum memories (QMs) are key elements of quantum information and communication systems, where precise control over both the storage and manipulation of information is essential for building scalable quantum networks \cite{kimble2008, Sangouard2011, heshami2016}. Atomic ensembles are suitable for implementing such memories, as they exhibit high optical depth, tunable frequency range and long-lived spin states, leading to high storage efficiency, extended storage durations, and robust phase coherence. Electromagnetically induced transparency (EIT) in a \(\Lambda-\)type three-level atomic system has become one of the most established methods to realize such memories \cite{Lei2023}, and has been implemented in cryogenic solid-state systems~\cite{Turukhin2001, Heinze2013}, ultracold and cold atomic ensembles~\cite{riedl2012, zhao2009, park2016,cho2016,vernaz2018, hsiao2018} and warm alkali vapors \cite{mair2002,katz2018,ma2022}. Notably, the phase of the retrieved signal can be directly measured and compared to that of the stored pulse \cite{chen2005,jeong2017}, thereby enabling the exploitation of this phase information to monitor additional parameters, such as the spatial displacement of the stored light, during the storage process. Such an ability is beneficial, as physical reconfiguration of atomic arrays can enhance quantum processing \cite{bluvstein2024}, and transport of quantum memories to improve quantum communication \cite{Gundogan2024} and astronomical interferometry \cite{bland2021} has been proposed.

The idea of storing light in a moving atomic medium and retrieving it at a different location has long been proposed~\cite{juzeliunas2003}. Here, the continuous wave (CW) control fields are launched perpendicular to the direction of the moving medium, and the passage of atoms through the optical fields define the adiabatic switching of the control field on and off. Such a configuration is helpful in cases that the arrival of the probe pulse cannot be fully determined, \emph{e.g.} it is originated in spontaneous processes. However, the other features of the storage process, the pulse profiles, storage time, etc., cannot be controlled. On the other had, it has been experimentally shown that spin coherence in warm alkali vapor can diffuse over extended spatial regions without dephasing \cite{zibrov2002, xiao2008}. However, in these studies, the spatial displacement of the stored information is not actively controlled and relies on the random motion of the atoms. Ginsberg \emph{et al.}~\cite{ginsberg2007} were the first to demonstrate the transfer of stopped light, storing an optical pulse in a Bose--Einstein condensate (BEC) and subsequently transferring it to another BEC located 160~\(\textup{\textmu}\)m away. Furthermore, the preservation of spin coherence during macroscopic displacement in an optical conveyor belt \cite{kuhr2003,xin2019} paved the way for transporting stored light in cold atoms confined by an optical lattice \cite{li2020}, where a stored optical pulse was displaced in a controlled manner through a hollow-core fiber over a distance \(\mathrm{\Delta x=  1.2}\)\,mm.

\begin{figure}[htbp]
    \includegraphics[width=0.9\linewidth]{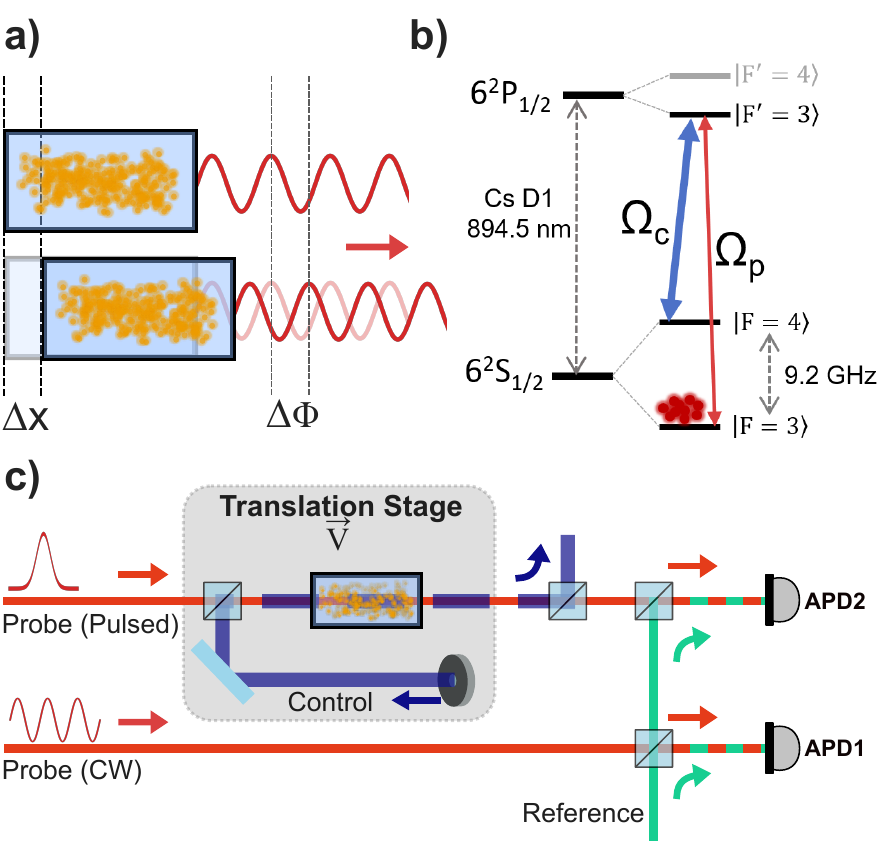}
    \caption{Basics of the measurement procedure. a) The basic principle behind measuring the displacement of the stored light. An optical pulse is stored in the medium and upon its retrieval, the phase of the retrieved pulse is fixed with respect to the memory frame. Displacement is measured by comparing phase of the retrieved pulse at two separate instances in the lab frame. b) The \(\Lambda\)-scheme used for realizing EIT and light storage. c) A schematic of the memory section. The cesium vapor cell is magnetically shielded and placed on the translation stage. All the optical fields are derived from one single laser and the control is launched from the translation stage to insure a fixed write/read phase as the stage moves. Polarization suppression clears control from probe, and before detection, the reference field is combined with the probe pulse sequence to generate a intra-pulse beating signal for detecting the phase shift. }
    \label{fig:Setup_Concept}
\end{figure}

In this Letter, we report on the controlled transport of an optical pulse stored within an alkali vapor cell. We furthermore infer information on the displacement of the stored pulse by monitoring its phase before and after the storage process. Figure \ref{fig:Setup_Concept} illustrates both the experimental setup and the fundamental measurement principle. The phase of the retrieved pulse is unchanged in the reference frame of the QM, but not in the lab frame. When the vapor cell is moved in space, comparing the phase of the retrieved pulse to that of a reference field leads to measuring the traveled distance. Light storage is accomplished via implementation of resonant EIT in the D1 line of cesium. The \(\Lambda\)-type atomic scheme consists of the ground states  \(\ket{\mathrm{F = 3}}\) and \(\ket{\mathrm{F = 4}}\), and the excited state \( \ket{\mathrm{F^\prime = 3}}\). A strong control field, offset-locked to the \(\ket{\mathrm{F} = 4} \text{--} \ket{\mathrm{F^\prime = 3}}\) transition, establishes the transparency condition. A small fraction of this control beam is frequency-shifted by 9.2\,GHz using an electro-optic modulator (EOM) to resonate with the \(\ket{\mathrm{F=3}} \text{--} \ket{\mathrm{F^\prime = 3}}\) transition, thus forming the probe beam. This method ensures an intrinsically high phase coherence between the probe and control fields. Pulse shaping is achieved through acousto-optic modulators (AOMs), driven by an arbitrary waveform generator (AWG). The probe and control beams have diameters of 1\,mm and 2\,mm, with peak powers \(\mathrm{P_P = 50\,\textup{\textmu W}}\) (\(\mathrm{\Omega_P = 2\pi \times7\,\/MHz}\)) and \(\mathrm{P_C = 10}\)\,mW (\(\mathrm{\Omega_C = 2\pi \times 70\,\/MHz}\)), respectively.

The vapor cell, containing 10\,Torr of nitrogen (\(\mathrm{N_2}\)) buffer gas, is housed within two layers of \(\textup{\textmu}\)-metal magnetic shielding and heated to 45\,\textdegree C to reach an optimal optical density. The entire cell assembly is mounted on a translation stage with a minimum incremental motion of 1\,nm. To precisely measure the phase shift of the probe pulse, we generate a reference field by passing the probe through an AOM, shifting its frequency by 80\,MHz. Prior to pulsing the probe beam and guiding it to the vapor cell, a portion of the probe beam is beated with this reference field and creates a CW 80\,MHz beat signal. This beat signal is then detected by an avalanche photodetector (APD1) and provides a stable reference for measuring the phase shift. This reference beat signal practically acts as a ruler to infer the displacement of the vapor cell during light storage. In order to monitor the phase shift in the storage process, a portion of the reference field is beated with the probe field after the vapor cell and prior to its detection by a second avalanche photodetector (APD2). This generates an intra-pulse 80\,MHz beat signal, which enables the accurate measurement of the phase evolution of the retrieved pulse relative to the input pulse during the motion of the translation stage. Both the CW reference beat and the probe intra-pulse beat signals can be viewed as being generated from two Mach-Zehnder interferometers (MZI) consisted of the probe field and the reference field; with the difference that for the latter, the probe beam is pulsed and goes through the vapor cell. More details on the laser assembly, pulse shaping and light storage is provided in the supplementary material. 

\begin{figure}
    \centering
    \includegraphics[width=0.95\linewidth]{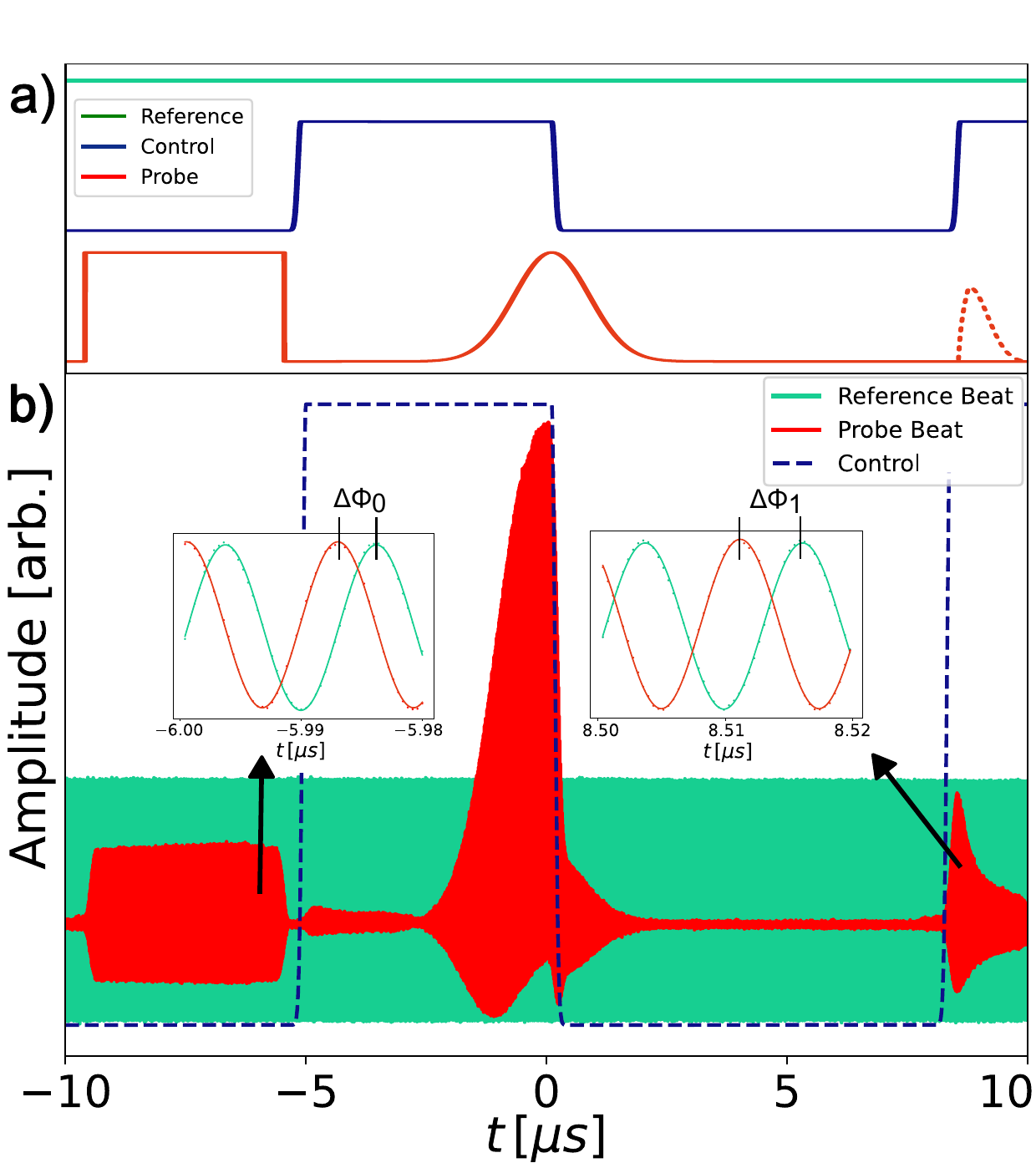}
    \caption{Light storage and phase detection. a) The respective pulse sequences. The dashed pulse in the probe panel illustrates retrieval of the stored pulse. b) An actual trace from the oscilloscope. The CW beating signal (green) acts as a ruler for measuring changes in phase. The probe pulse sequence (red) consists of a square pulse (as baseline phase) and a Gaussian pulse (to be stored). The insets show 20\,ns long traces at \(\mathrm{t = -6 \, \textup{\textmu} s}\) (to measure \(\mathrm{\Delta \Phi_0}\)) and \(\mathrm{t = +8.5 \, \textup{\textmu}s}\) (to measure \(\mathrm{\Delta \Phi_1}\)).}
    \label{fig:pulse_seq}
\end{figure} 

Figure \ref{fig:pulse_seq}a) shows the pulse sequence of the storage process and Figure \ref{fig:pulse_seq}b) shows an actual trace of the detected light fields on the oscilloscope. The green background curve shows the reference beat signal, the red curve demonstrates the detected probe pulse, carrying an intra-pulse beating, and the dashed line indicates the intensity profile of the control field. The control field is turned on at \(t=-5\,\textup{\textmu s}\) (pumping atoms to \(\ket{\mathrm{F=3}}\) and preparing the EIT conditions) and is adiabatically switched off at \(\mathrm{t=0\,s}\). After a storage time \(\mathrm{\tau_s}  = 8.5\,\textup{\textmu s}\), it is switched back on again to retrieve the stored pulse from the atomic ensemble. Prior to activating the control field, the probe is switched on for a short duration. The phase of the probe field may fluctuate from shot to shot due to mechanical vibrations and the laser's finite coherence length; therefore, this initial pulse is essential and provides us with a baseline phase difference \(\Delta \Phi_0\) between the background beat signal and the probe field. It is important to note that \(\Delta \Phi_0\) is measured in the absence of EIT, where light-drag effects \cite{fizeau1851} are negligible. After the control being on for 2.5 \(\textup{\textmu s}\), a 2 \(\textup{\textmu s}\) Gaussian-shaped probe pulse is sent into the vapor cell. Its timing is aligned with the onset of the control field's switch-off, ensuring optimal overlap for maximum storage efficiency.

After retrieval of the stored pulse, the phase of the retrieved pulse is compared with that of the background beat and labeled as \(\Delta \Phi_1\). We can define the phase evolution of the probe field (P) during storage process as
\begin{equation}
   \mathrm{\Delta \Phi_{P}^{R/M} \equiv \Delta \Phi_1^{R/M} - \Delta \Phi_0^{R/M}} \label{eq:phase_evol}
\end{equation}
with R and M indicating the cases when the stage is at "Rest" or in "Motion", respectively. Now, if we measure a \(\mathrm{\Delta \Phi_{P}^R}\) and subsequently a \(\mathrm{\Delta \Phi_{P}^M}\) for some storage time \(\mathrm{\tau_s}\), we can define the phase shift due to this translation (Tr.) as:
\begin{equation}
    \mathrm{\Delta \Phi_{Tr.}  \equiv \Delta \Phi_{P}^M - \Delta \Phi_{P}^R} = -\vec{\mathrm{k}}_{\mathrm{P}}. \Delta \vec{\mathrm{x}} = -\vec{\mathrm{k}}_{\mathrm{P}}. \vec{\mathrm{V}} \mathrm{\tau_s}
    \label{eq:Dphi_vs_V}
\end{equation}

\noindent
where \(\vec{V}\) is the average velocity of the translation stage during storage, resulting in a displacement of \(\Delta \vec{\mathrm{x}} = \vec{\mathrm{V}} \mathrm{\tau_s}\) for the vapor cell \(-\) and thus the stored pulse.

\begin{figure}
    \centering
    \includegraphics[width=0.95\linewidth]{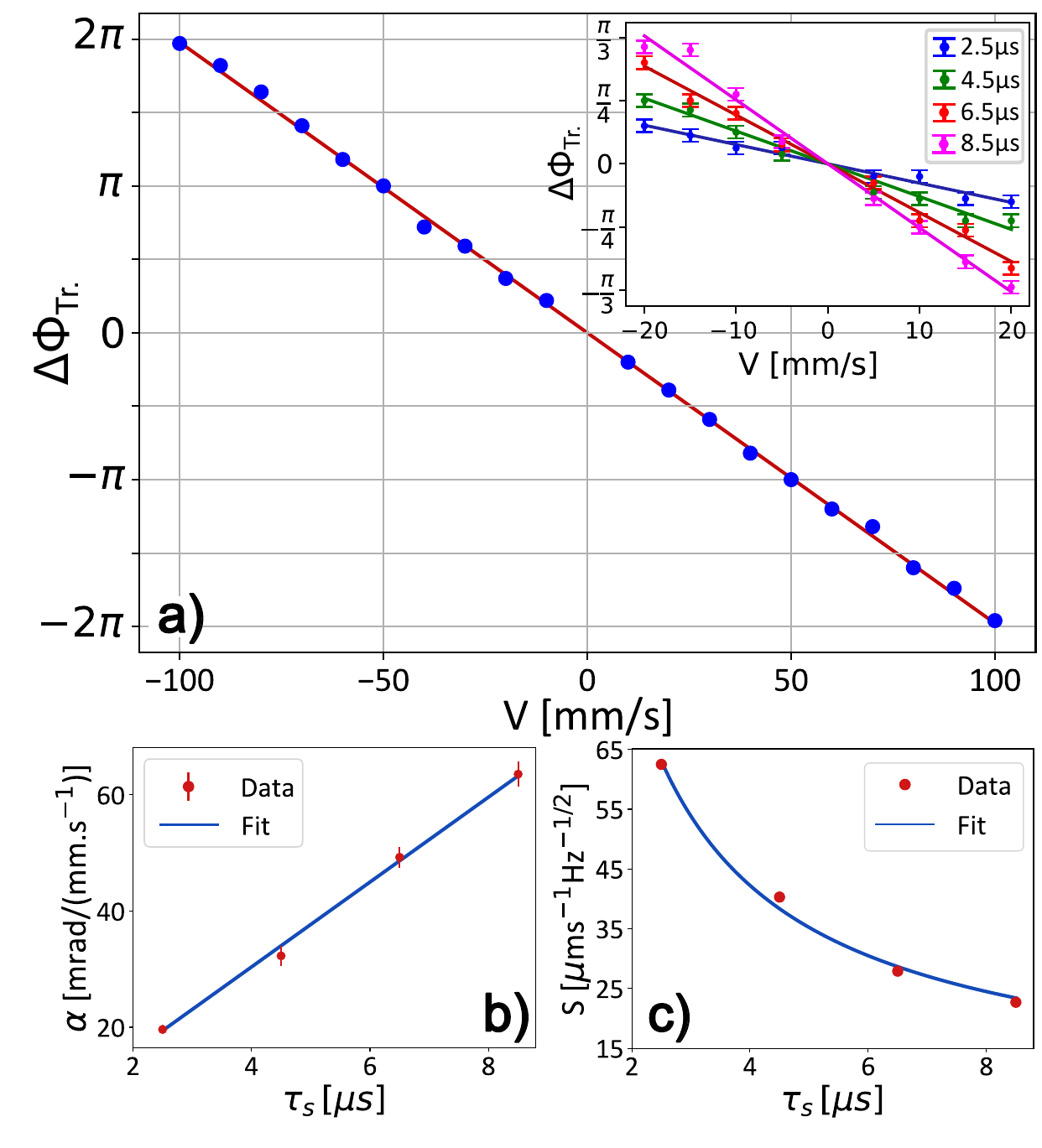}
    \caption{Spatial transport of stored light. a) Measured phase shift vs. stage speed, for \( \mathrm{\tau_s = 8.5\,\textup{\textmu s}}\). The stored pulse is translated a full wavelength on either direction with \(\mathrm{\Delta\Phi_{Tr.} = \pm 2\pi}\). The inset shows the same measurement for different storage times. b) Plot of coefficient \(\mathrm{\alpha}\) vs. \(\mathrm{\tau_s}\). The fitted line has a slope \(\mathrm{m = 7.31 \pm 0.33}\)\,mrad/nm, which matches the expected \(\mathrm{k_P = 7.02}\)\,mrad/nm. c) Calculated sensitivity of the setup when used as a velocimeter.}
    \label{fig:phase shift vs. speed}
\end{figure}

Figure \ref{fig:phase shift vs. speed}a shows the resulting phase shift for different velocities, from -100\,mm/s to +100\,mm/s, with \(\tau_s = 8.5\,\textup{\textmu s}\). As expected, the measured phase shift exhibits a linear relation to V, \emph{i.e.}, \(\mathrm{\Delta \Phi_{Tr.} = -\alpha V}\), with \(\mathrm{\alpha = 63.6\,(\pm2.2)}\)\,mrad/(mm.s\(^{-1}\)). Observing phase shifts up to \(\pm 2\pi\) means that the stored pulse has been displaced by \(\mathrm{\Delta x = \lambda_P}\) on either direction, with \(\mathrm{\lambda_P}\) being the wavelength of the probe field. Moreover, it is possible to track the displacement of the stored pulse anywhere between these two extremes. 

Even though derived from different physical principles, the phase shift observed in our setup, Eq.\ref{eq:Dphi_vs_V}, follows the same mathematical description as that of EIT-enhanced light-drag \cite{kuan2016,chen2020, solomons2020, banerjee2022}, with the delay time \(\mathrm{\tau_d}\) replaced with storage time \(\mathrm{\tau_s}\). Hence, in a similar manner, this relation can be exploited to create a velocimeter. In order to fully verify the validity of Eq.\ref{eq:Dphi_vs_V} in our setup, we have repeated this measurement for other storage times, \(\tau_s = [2.5\,\textup{\textmu s},\ 4.5\,\textup{\textmu s},\ 6.5\,\textup{\textmu s},\ 8.5\,\textup{\textmu s}]\), with stage velocities in the range (-20\,mm/s, +20\,mm/s). The corresponding results are presented as the inset of Fig.\ref{fig:phase shift vs. speed}a, which shows clear increase in the steepness of the plots as the storage time increases. Fig. \ref{fig:phase shift vs. speed}b shows the plot of \(\alpha\) vs. \(\tau_s\), following a linear trend with the slope \(\mathrm{m=7.31\,(\pm 0.33)}\)\,mrad/nm, covering the actual wave number of the probe \(\mathrm{k_P = 7.02}\)\,mrad/nm. In each measurement, the oscilloscope trace is averaged 16 times and the phase shift is averaged over 20 runs. For the case of \(\tau_s = 8.5\,\textup{\textmu s}\), this results in a total integration time \(T = 6.9\)\,ms. The calculated standard error of the phase shift \(\hat{\sigma}_{\Phi} = 17.4\)\,mrad, equivalent to a velocity resolution of \(\mathrm{\Delta V \approx 270} \, \textup{\textmu}\)m/s. This translates to a short-term sensitivity (S) of 22.7 \textup{\textmu m}\,s\,\(^{-1}\)Hz\(^{-1/2}\). Fig. \ref{fig:phase shift vs. speed}c shows the plot of S vs. \(\tau_s\). The fitted line follows the equation \(\mathrm{S = \frac{\hat{\sigma}_{\Phi}}{k_P \tau_s} \sqrt{T}}\), where integration time \(\mathrm{T = \beta (\tau_s + \epsilon)}\), with \(\mathrm{\beta}\) representing the total number of pulse sequences considered in each data point (oscilloscope averaging \(\times\) measurement runs), and \(\mathrm{\epsilon}\) representing length of the pulse sequence excluding the storage duration. As the storage time is increased, the sensitivity also improves, following a \(1/\mathrm{\sqrt{\tau_s}}\) trend. 

The phase fluctuation in our setup is 4 orders of magnitude worse than the ultimate photon shot-noise limit, 17.4\,mrad vs. 8.3\,\(\textup{\textmu}\)rad. However, our achieved resolution in velocity measurement is comparable to that of the usual laser doppler velocimeters \cite{norgia2017, sun2018, zhang2019} and EIT-enhanced longitudinal light-dragg measurements \cite{kuan2016, chen2020}. It is worth to mention that our acquired values are not the ultimate performance of this scheme, since extending the storage time and improving the phase stability of the interferometer can significantly increase the velocimetry resolution. However, the current experimental setup imposes certain limitations on further enhancing the sensitivity. Presently, the storage time is limited to \(8.5 \, \textup{\textmu s}\). Extending this duration would require improved magnetic shielding, since the translation stage is highly magnetic. Furthermore, the laser used has a short-term  linewidth of \(\mathrm \sim 100\)\,kHz, which limits its temporal coherence to about \(10\,\textup{\textmu s}\). As a result, the temporal separation between the baseline and retrieved pulses should not significantly exceed \(10\,\textup{\textmu s}\); otherwise, the measured phases (\(\Delta \Phi_0\) and \(\Delta \Phi_1\)) would no longer be reliably comparable. Nevertheless, reducing the natural linewidth of the laser to below 1\,kHz is well established \cite{fu2017}, thereby extending the laser’s coherence time to the millisecond range. This time scale fits well with the storage times achievable with vapor cells in compact and relatively simple memory setups \cite{wang2022}. Since the beating signal is generated via a MZI without any active stabilization, the phase relation between the probe and reference fields fluctuates from shot to shot, and averaging the trace of the oscilloscope beyond 16 times reduces the actual phase stability. At this limit, measurement bandwidth reaches the usual frequency associated to ambient mechanical vibrations ( \(\mathrm \sim 100\)\,Hz). Therefore, active phase stabilization of the reference beat note with respect to the probe beat note is a necessary step moving forward. As an estimate, by stabilizing the phase drift of the MZI to below 1 degree \cite{mivcuda2014} and extending the storage to 1\,ms, our sensitivity could be enhanced by three orders of magnitude to \(\mathrm{S = 20\, nm\,s^{-1} \, Hz^{-1/2}}\).

In summary, we have reported on spatial displacement of a stored pulse in a warm alkali vapor cell. Moreover, by monitoring the phase of the retrieved pulse, we were able to accurately track the displacement of the stored pulse up to one optical wavelength along both \(\pm \vec{\mathrm{z}}\) directions. This method can be utilized as an effective motion sensor, as explored in our previous work \cite{ahmadi2025}. The findings open new avenues for more complex implementations of quantum memories in warm vapors and offer applications beyond traditional domains such as quantum information and communications.

We acknowledge support from the Deutsche Forschungsgemeinschaft (DFG) under project number 448245255, and from the German Aerospace Center (DLR) with funding provided by the Federal Ministry for Economic Affairs and Climate Action (BMWK) under Grant No. 50WM2347 (OPTIMO-III). We also gratefully acknowledge support from the National Science and Technology Council (NSTC) of Taiwan under Grant No. 109-2923-M-006-003-MY3.

\bibliography{Refs}

\end{document}

%\title{Velocimetry with stopped light at room temperature}
\title{Supplemental Material:\\Controlled Displacement of Stored Light at Room Temperature}

\author{Arash Ahmadi}
\email{arash.ahmadi@physik.hu-berlin.de}
% \affiliation{Institut für Physik and Center for the Science of Materials Berlin (CSMB), Humboldt-Universität zu Berlin, Newtonstr. 15, Berlin 12489, Germany}

\author{Ya{\u{g}}{\i}z Murat}
% \affiliation{Institut für Physik and Center for the Science of Materials Berlin (CSMB), Humboldt-Universität zu Berlin, Newtonstr. 15, Berlin 12489, Germany}

\author{Pei-Chen Kuan}
% \affiliation{Department of Physics, National Cheng Kung University, Tainan 70101, Taiwan}

\author{Mustafa Gündo{\u{g}}an}
\email{mustafa.guendogan@physik.hu-berlin.de}
% \affiliation{Institut für Physik and Center for the Science of Materials Berlin (CSMB), Humboldt-Universität zu Berlin, Newtonstr. 15, Berlin 12489, Germany}

\author{Markus Krutzik}
% \affiliation{Institut für Physik and Center for the Science of Materials Berlin (CSMB), Humboldt-Universität zu Berlin, Newtonstr. 15, Berlin 12489, Germany}
% \affiliation{Ferdinand-Braun-Institut (FBH), Gustav-Kirchoff-Str. 4, Berlin 12489, Germany}

\date{\today}

\maketitle

\section{I. Experimental Setup}

\begin{figure}[H]
    \centering
    %YM: I put the png version on purpose because my pc cannot easily render the pdf version 
    \includegraphics[width=0.6\linewidth]{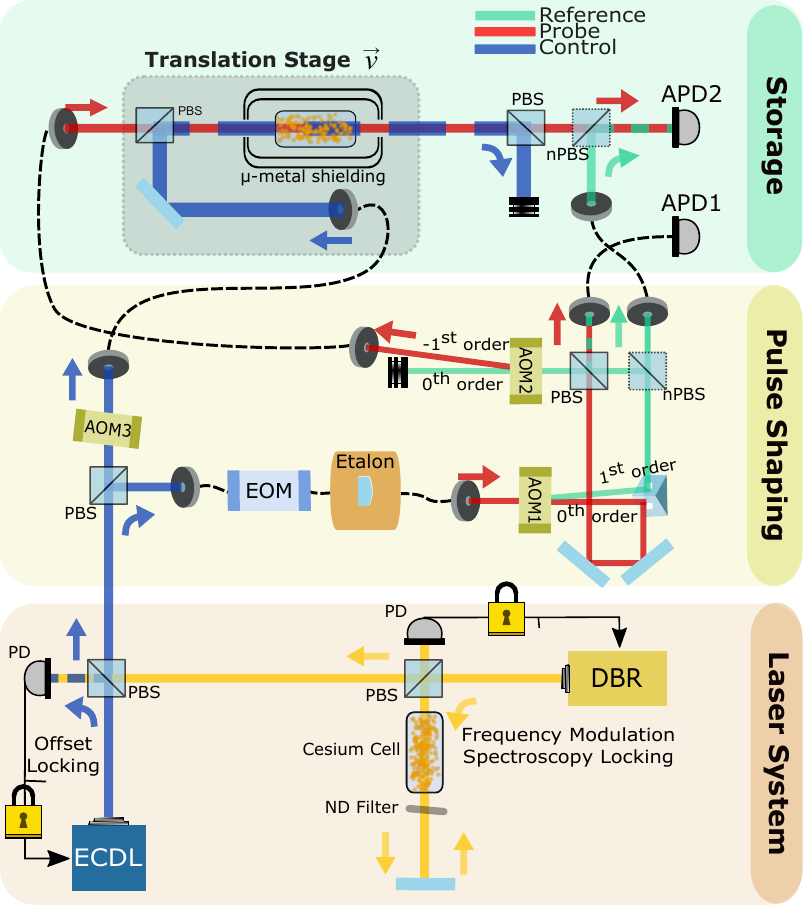}
    \caption{Schematic overview of the experimental setup. The setup consists of three main sections: (i) Laser system, featuring frequency-stabilized DBR master laser, and an ECDL slave laser; (ii) Pulse shaping module, where control, probe, and reference fields required for the operation of the setup are created and sequenced by using AOMs, an EOM, and a filter etalon; (iii) Storage module, containing a cesium vapor cell—filled with 10 Torr nitrogen buffer gas—sits on a motorized translation stage inside two layer of magnetic shielding, maintained at 45\,\textdegree C.  PBS: polarizing beam splitter; nPBS: non-polarizing beam splitter; APD: avalanche photodiode; PD: photodetector; ND: neutral density filter. Reference (green), probe (red), and control (blue) beam paths are indicated. Black dashed lines represent optical fibers.}
    \label{fig:full_setup}
\end{figure}

The experimental setup can be divided into three main sections, namely, the laser system, pulse shaping, and storage modules (see Fig. \ref{fig:full_setup}). These sections are discussed in that order in the following paragraphs.

The setup incorporates two main laser sources: a distributed Bragg reflector (DBR) laser as the master laser, and an extended cavity diode laser (ECDL) as the slave laser. The DBR laser is locked to the  \(\ket{\mathrm{F=4}} \text{--} \ket{\mathrm{F^\prime = 4}}\) hyperfine states of cesium D1 line, using a reference cell. This locking is done by means of frequency modulation spectroscopy. Subsequently, the ECDL laser is locked to \(\ket{\mathrm{F=4}} \text{--} \ket{\mathrm{F^\prime = 3}}\) transition. For both locking schemes, a Red Pitaya development board incorporating Linien locking software \cite{Wiegand2022} is utilized, along with an in-house phase-frequency detector (PFD) for the offset locking. After locking, the standard deviation of the lasers' frequencies are on the order of 200\,kHz. All the necessary fields for the storage, retrieval and phase measurement are derived from the ECDL laser which has a linewidth of <100\,kHz.

After locking the lasers to relevant transitions, the ECDL laser is sent to the pulse shaping module to create and sequence the necessary pulses. One part of the ECDL is sent through an acousto-optic modulator (AOM) denoted as AOM3 operated at 80\,MHz. This portion of the ECDL constitutes the control field (blue in Fig. \ref{fig:full_setup}). The other part of the ECDL laser is then sent to a phase electro-optic modulator (EOM). The EOM is modulated at 9.2\,GHz to match the ground state hyperfine splitting of cesium, and the +1st order sideband is isolated by a monolithic, temperature-controlled etalon filter adapted from \cite{Heller2023}. This filter has an extinction ratio of 43\,dB, and is temperature stabilized with stability \( \mathrm{\Delta T < 1} \)\,mK by a Meerstetter thermoelectric cooler (TEC) controller. The isolated sideband is then sent to a continuously driven AOM (AOM1) and an 80\,MHz optical reference beat is created by bringing together the 0th and +1st orders and sent to an avelanch photodetector (APD1) in the storage module (the green curve in Fig. 2.b of the main text). The +1st order constitutes the reference field (green in Fig. \ref{fig:full_setup}). One portion of this reference field is then sent to another AOM (AOM2), where -1st order constitutes the probe field (red in Fig. \ref{fig:full_setup}). The other portion of the reference field is directly sent to the storage module to create an optical beat with the probe field pulse sequence after the storage process. All of the RF generators used in the control electronics are clocked to a common 100\,MHz reference crystal to prevent any phase instability between the electronics.

At the storage module, probe and control beams are overlapped with orthogonal linear polarizations, H and V respectively. The control field is filtered out by spatial and polarization filtering with an extinction ratio of 67\,dB. Subsequently, the reference field from the pulse shaping module is superposed with the retrieved probe field using a non-polarizing beam splitter (NPBS) to generate another inter-pulse 80\,MHz beat note at APD2 (the red curve in Fig. 2.b of the main text). The reference beat created by AOM1 is also detected simultaneously at APD1. The cesium cell used to store the probe field is around 10\,mm wide and 35\,mm long with anti-reflection coated windows attached at 11\,\textdegree, and it includes a buffer gas of 10 Torr nitrogen. The cesium cell is mounted on a motorized linear translation stage (Newport XMS-50) which has a maximum speed of 300\,mm/s, a travel range of 50\,mm and a minimum incremental motion of 1\,nm. The cell is shielded by two 1\,mm thick layers of \(\textup{\textmu}\)-metal. These layers have outer diameters of 40 and 60\,mm (to maximize the shielding effect \cite{dubbers1986}) with a length of 200\,mm. The cell is placed in by using an in-house aluminum cell holder. The cell holder is heated by ceramic resistive heaters and stabilized to 45\,(\(\pm0.005\))\,\textdegree C by another Meerstetter TEC controller.

\section{II. Phase Detection and Sensitivity}
\begin{figure}[H]
    \centering
    \includegraphics[width=0.5\linewidth]{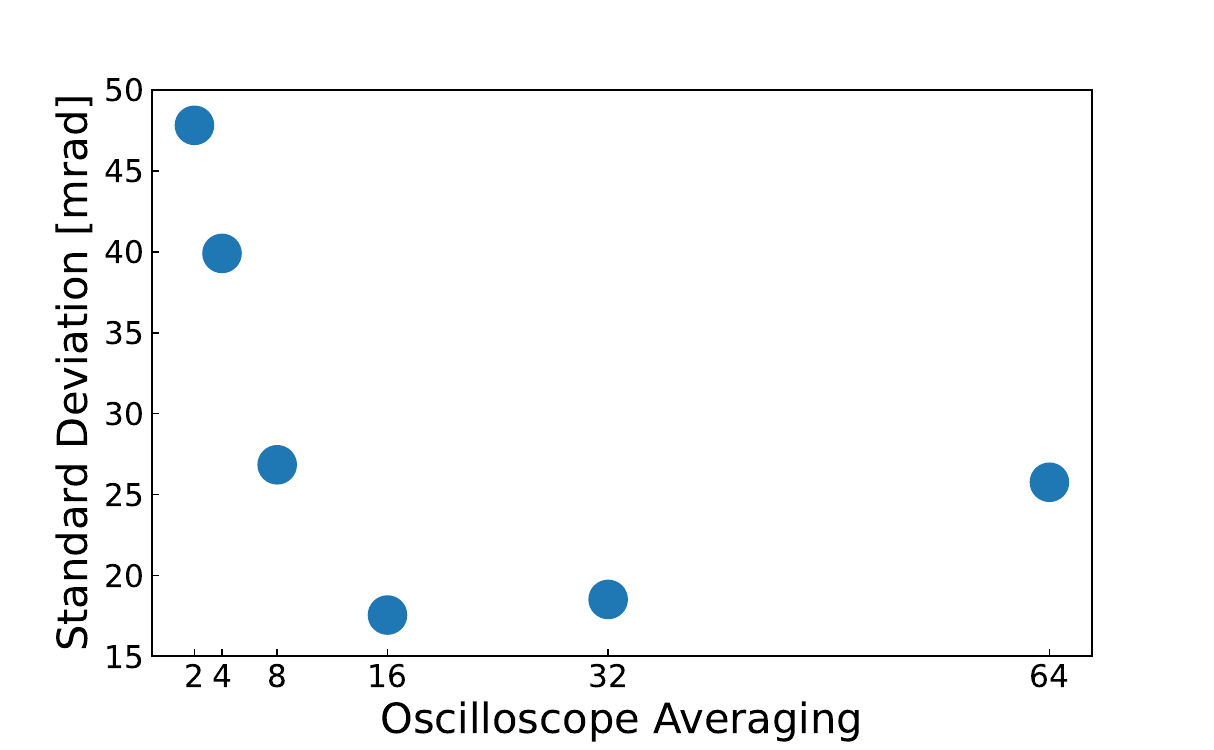}
    \caption{The standard deviation of phase difference measurements between the two beat notes recorded at APD1 and APD2. Each point is recorded 20 times.}
    \label{fig:osc_avg}
\end{figure}

As can be seen from Fig. \ref{fig:full_setup}, two separate Mach-Zehnder type interferometers (MZI) lead to formation of two beat signals, a CW one  acting as a reference beat and the intra-pulse beat note of the probe pulse sequence. The two beat signals are detected by APD1 and APD2 respectively. The detected signals are sent to an oscilloscope (Tektronix MDO34) to measure their phase relation. For each data point, an averaged trace of 1 mega samples (MS) at 2.5 GS/s resolution (\(\mathrm{\Delta t = 400 \,\textup{\textmu} s}\)) was recorded for 20 times. Out of the full recorded trace, the trace of one pulse sequence (\(\mathrm{\Delta t = 20 \,\textup{\textmu} s}\)) was separated and analyzed for phase relations. The standard phase error for different averaging settings of the oscilloscope was measured. The results are presented in Fig. \ref{fig:osc_avg}. The most optimal oscilloscope averaging is found around 16 to 32 averages. Taking the sampling frequency and the recorded samples into account, the resultant integration times correspond to a frequency range of 80-150\,Hz, which agrees with the environmental ground vibrations present in urban environments \cite{ainalis2018}. When averaging beyond 32 times, these environmental ground vibrations contribute additional variance between the two MZIs, causing the observed rise in the error.

\section{III. Memory Characterization}

\begin{figure}[h]
    \centering
    \includegraphics[width=0.5\linewidth]{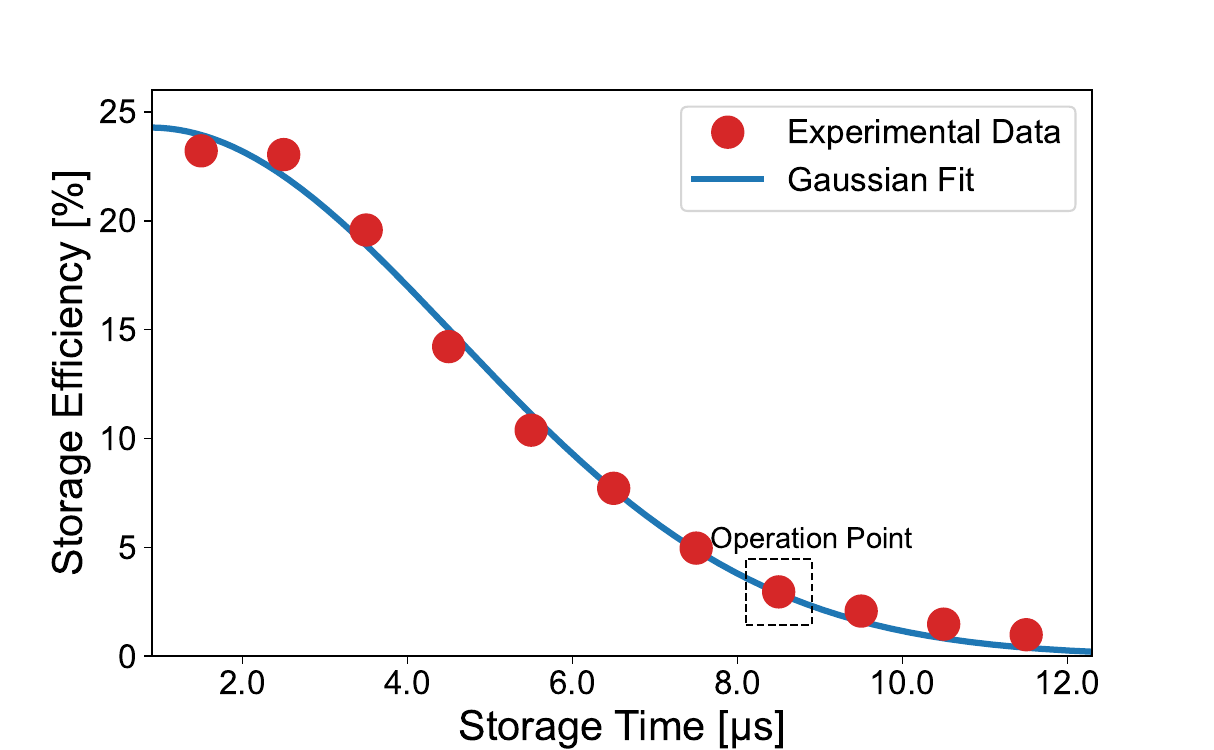}
    \caption{Storage efficiency as a function of storage time. The operation point is at 8.5\,\(\mu\)s storage with approximately \(3\%\) efficiency, and was used for the main results presented in the paper.}
    \label{fig:memory_characterization}
\end{figure}

In order to extend the maximum displacement of the stored light as well as the sensing resolution, storage efficiency and storage time are the parameters of main concern for the purposes of this work. Probe field is a Gaussian with a 2\,\(\mu\)s FWHM, and the control field has half-Gaussian rise and fall profile with 200\,ns FWHM for optimized storage and retrieval. In order to obtain the storage efficiency, with a slight modification to the optical setup, the probe pulse is split 50/50; one part is detected by APD1 before entering the vapor cell and serves as a reference to calibrate APD2 relative to APD1. During storage measurements, the intensity profile of the retrieved pulse is compared to that of the reference probe pulse, allowing the memory efficiency to be inferred. The storage efficiency vs. storage time is plotted in Fig.~\ref{fig:memory_characterization}. By measuring electromagnetically induced transparency (EIT) linewidth vs. control power,  the pure ground state decoherence of the system is estimated to be approximately \(20\)\,kHz, which sets an upper limit of 50\,\(\mu\)s in the memory lifetime. However, due to the other decoherence and broadening mechanisms (such as diffusion, Doppler broadening in the hyperfine manifold, etc.) present in the system, the efficiency decay follows a Gaussian with a time scale limited to \(\sim4.5\,\mu\)s and a zero storage time efficiency of 24\%.

%Bibliography
%\bibliographystyle{unsrt}  
\bibliography{Ref_Sup}